\documentclass[aps,prl,showpacs,superscriptaddress,twocolumn,longbibliography]{revtex4}

\usepackage{graphicx}
\usepackage{dcolumn}
\usepackage{amsmath}
\usepackage{epsfig}
\usepackage{bm}
\usepackage{amssymb}
\usepackage{float}
\usepackage{natbib}
\usepackage{color}
\usepackage{hyperref}
\hypersetup{colorlinks=true,linkcolor=blue,citecolor=blue,urlcolor=black}

\newcommand{\ket}[1]{| #1 \rangle}
\newcommand{\supp}{\cite{Supplementarymaterial}}
\DeclareMathOperator{\erf}{erf}

\begin{document}

\title{Quantum Rabi oscillations in coherent and in mesoscopic ``cat" field states}

\author{F. Assemat}
\author{D. Grosso}
\author{A. Signoles}
\author{A. Facon}
\author{I. Dotsenko}
\author{S. Haroche}
\author{J.M. Raimond}
\author{M. Brune}
\author{S. Gleyzes}
\affiliation{Laboratoire Kastler Brossel, Coll\`ege de France,
  CNRS, ENS-Universit\'e PSL,
  Sorbonne Universit\'e, \\11, place Marcelin Berthelot, 75005 Paris, France}
\email{gleyzes@lkb.ens.fr}

\date{\today}

\begin{abstract}
The simple resonant Rabi oscillation of a two-level system in a single-mode coherent field reveals complex features at the mesoscopic scale, with oscillation collapses and revivals. Using slow circular Rydberg atoms interacting with a superconducting microwave cavity, we explore this phenomenon in an unprecedented range of interaction times and photon numbers. We demonstrate the efficient production of `cat' states, quantum superposition of coherent components with nearly opposite phases and sizes in the range of few tens of photons. We measure cuts of their Wigner functions revealing their quantum coherence and observe their fast decoherence. This experiment opens promising perspectives for the rapid generation and manipulation of non-classical states in cavity and circuit Quantum Electrodynamics.
\end{abstract}

\maketitle

The Rabi oscillations of a two-level atom in a resonant, single-mode coherent field state is one of the simplest phenomena in quantum optics.  Nevertheless, it exhibits surprisingly complex features at the mesoscopic scale (few tens of photons)~\cite{EberlyPeriodicSpontaneousCollapse1980,FleischhauerRevivalsMadeSimple1993,RaimondManipulatingquantumentanglement2001,FeranchukAnalyticalanalysiscollapserevival2009}. The oscillations, at an angular frequency $\Omega_0\sqrt{\overline n}$,   collapse and revive ($\overline n$ is the average photon number in the coherent state; $\Omega_0$ is the vacuum Rabi frequency measuring the atom-field coupling). The collapse, occurring on a time scale $T_c=2\sqrt{2}/\Omega_0$, results from the quantum field amplitude uncertainty and from the corresponding dephasing of the Rabi oscillations. The (first) revival, around $T_r=4\pi\sqrt{\overline n}/\Omega_0$, results from the rephasing of oscillations associated to different photon numbers~\supp. This revival provides a landmark illustration of field amplitude quantization~\cite{BruneQuantumRabiOscillation1996}. Between collapse and revival, the field evolves into an entangled atom-field state, involving  two coherent states with different phases~\cite{EiseltCalculationquasiprobabilitiesdamped1989,Gea-BanaclocheCollapserevivalstate1990,Gea-BanaclocheAtomfieldevolution1991,BuzekSchrodingerCatStatesResonant1992,AverbukhFractionalrevivalsJaynesCummings1992}. It is called a ``cat state" in memory of Schr\"odinger's metaphor. Close to $t=T_r/2$, the atomic state factors out of a field ``cat", superposition of coherent states with opposite phases~\supp.

These phenomena can be observed in systems implementing the Jaynes and Cummings model, a spin-$1/2$ coupled to a one-dimensional harmonic oscillator~\cite{JaynesComparisonquantumsemiclassical1963}. Ions in traps~\cite{MeekhofGenerationnonclassicalmotional1996,WinelandNobelLectureSuperposition2013}, cavity quantum electrodynamics (CQED)~\cite{BruneQuantumRabiOscillation1996,RaimondManipulatingquantumentanglement2001} and circuit quantum electrodynamics (cQED)~\cite{VlastakisDeterministicallyEncodingQuantum2013,WangSchrodingercatliving2016} are thus ideal platforms for this observation. 

Nevertheless, revival observations have so far been limited to small photon numbers since experiments face formidable challenges. 
For microwave CQED with superconducting cavities crossed by fast Rydberg atoms, the interaction time is limited to a few vacuum Rabi periods, $2\pi/\Omega_0$. Revivals have been observed only for $\overline{n}\simeq1$~\cite{RempeObservationquantumcollapse1987,BruneQuantumRabiOscillation1996}. Early revivals induced by a time-reversal of the collapse can be observed for larger $\overline n$ values  (about 10), but the maximum separation of the cat components is small~\cite{AuffevesEntanglementmesoscopicfield2003,MeunierRabioscillationsrevival2005}. Ion traps have similar limitations~\cite{MeekhofGenerationnonclassicalmotional1996,LvReconstructionJaynesCummingsfield2017}. In cQED, the limited coherence time of tunable superconducting qubits makes it difficult to observe long-term dynamics in the resonant regime~\cite{JohanssonVacuumRabiOscillations2006}. Large cat-state preparation so far relies instead on the dispersive, non-resonant interaction~\cite{KirchmairObservationquantumstate2013,VlastakisDeterministicallyEncodingQuantum2013}, in which the atom is simply a transparent dielectric material with a state-dependent index of refraction~\cite{RaimondManipulatingquantumentanglement2001}.

In this Letter, we push the quantum revival phenomenon at a much larger scale. Using slow circular Rydberg atoms crossing a high-$Q$ superconducting cavity, we achieve atom-field interaction times up to twenty vacuum Rabi oscillations periods. We observe the complete first revival for $\overline{n}=13.2$.  Resetting the atom close to $t=T_r/2$, we leave in the cavity a cat state. We observe the quantum revivals in this initial cat state. We use them to measure the cat state Wigner function and to investigate its  fast decoherence.

\begin{figure}
\includegraphics[width=8.5cm]{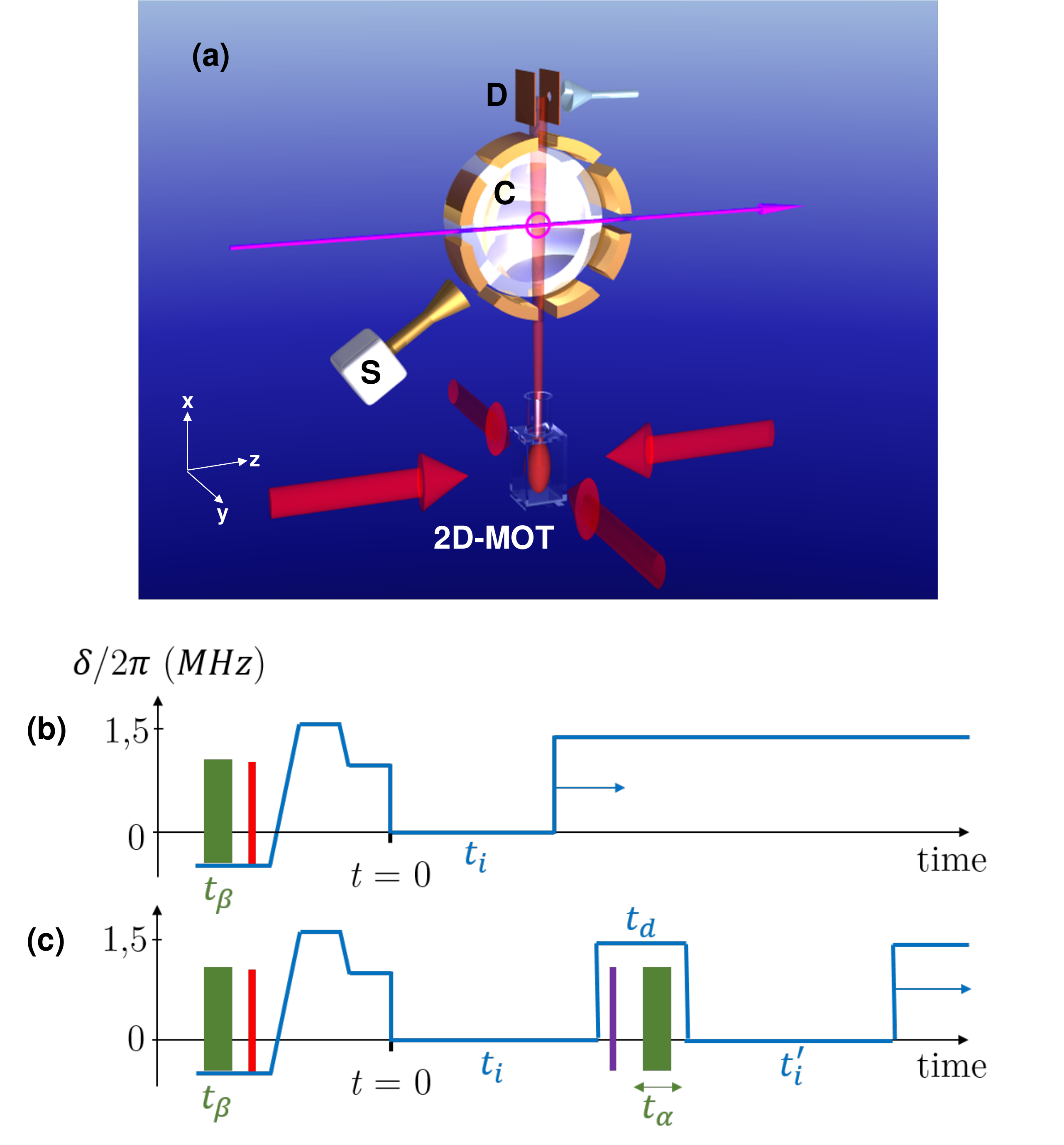}
\caption{(a) Sketch of the experimental set-up (only one cavity mirror shown for clarity), with the axis conventions. (b) Timing of the Rabi oscillation sequence. The solid blue line depicts the atom-cavity detuning $\delta$ as a function of time (time scale is qualitative). The resonant interaction starts at $t=0$ and lasts for a variable interaction time $t_i$. The green bar depicts the mw injection in $C$ starting at $t=-71\ \mu$s and lasting $t_\beta$, the red bar the exciting laser pulse at $t=-25\ \mu$s. The circular state preparation sequence takes place between the laser pulse and $t=0$.  (c) Timing of the cat revival experiment. Same conventions as for (b). The two resonant interactions last for a fixed time $t_i = 60\ \mu$s and a variable time $t'_i$. They are separated by the variable delay $t_d$ during which $\ket g$ is eliminated (purple bar) and an injection lasting $t_\alpha$ is performed (green bar).}
\label{fig:scheme} 
\end{figure}

The experiment is sketched in Fig.~\ref{fig:scheme}(a). Additional details are given in~\supp. A 2D-MOT and an additional longitudinal velocity selection produce a Rubidium atomic beam propagating upwards along the $Ox$ axis at an average velocity $v=8.4$~m/s towards the cavity $C$.

Inside the cavity, atomic samples are selectively prepared in $\ket e$, the circular state $51C$ with a principal quantum number 51, by laser, radio-frequency and microwave excitation~\cite{SignolesCoherentTransferLowAngularMomentum2017}.  Each sample contains 0.08 atoms on the average. Events with two atoms simultaneously present in $C$ have thus negligible influence. The cavity is tuned close to resonance with the  $\ket e\rightarrow \ket g$ transition at 51.1~GHz ($\ket g$ is the $50C$ circular state). An electric field along the cavity axis, $Oy$, produced by a voltage applied across the mirrors, stabilizes the circular states and makes it possible to Stark-tune the atomic transition frequency  in or out of resonance with $C$. After the interaction with the cavity mode, the atoms drift towards a field-ionization counter $D$, allowing us to measure the populations in $\ket e$ or $\ket g$ (see~\supp\ for details).

The cavity $C$, cooled down to 1.5~K by a wet $^4$He cryostat, sustains a linearly polarized  Gaussian standing-wave mode with a waist $w=6$~mm~\cite{RaimondManipulatingquantumentanglement2001}. Its damping time is $T_{Cav}=8.1\pm0.3$~ms. Its temperature corresponds to an average thermal photon number $n_{th}=0.38$. In order to reduce the residual photon number in $C$, we send absorbing atoms prepared in $\ket g$, starting 2.7~ms before the sequence. A microwave source $S$, coupled to $C$, performs tunable coherent displacements of its mode. The injected amplitude $\beta$ is a linear function of the injection time $t_\beta$. For  injections lasting more than $\approx 100$~ns, the calibration is $|\beta|=0.257( t_\beta (\mu s)-0.05)$~\supp.

We first record the vacuum Rabi oscillations with no injection ($\beta=0$) and an atom initially in $\ket e$. The sequence timing is sketched in  Fig.~\ref{fig:scheme}(b). The initial state $\ket e$ is prepared in a large electric field, resulting in a  $\delta/2\pi=1.4$~MHz atom-cavity detuning, for which the interaction is negligible. We abruptly set $\delta=0$ at $t=0$. After a variable interaction time $t_i$, we set back $\delta$ to a  large value ($2\pi\times 4.04$~MHz), halting the evolution. We measure the probability, $P_g$, for finding the atom in $\ket g$. 

\begin{figure}
\includegraphics[width=8.5cm]{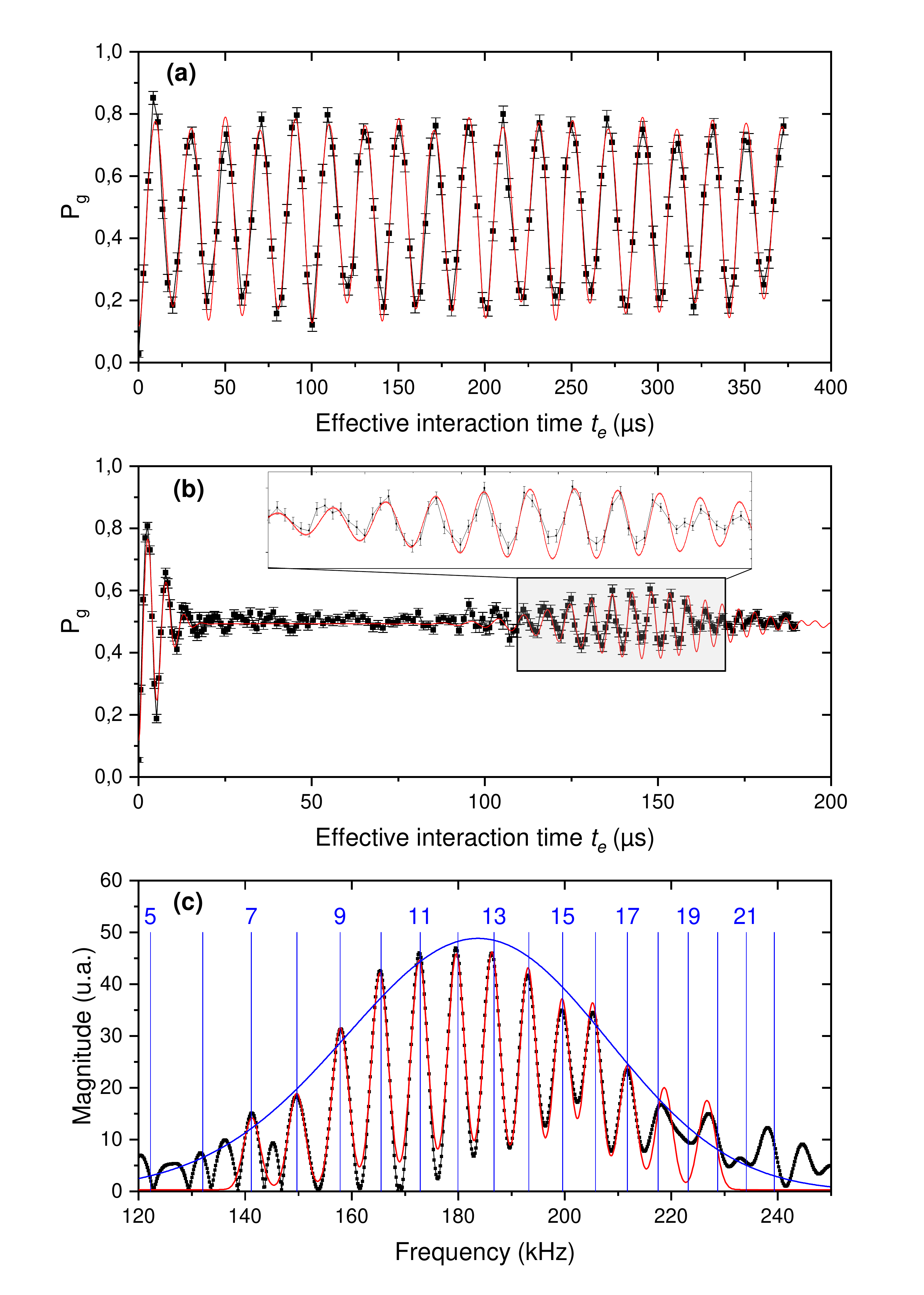}
\caption{(a) Vacuum Rabi oscillation. Dots with statistical error bars joined by a thin black line: experimental probability $P_g(t_e)$ as a function of the effective interaction time $t_e$. Solid red line: numerical simulation of the experiment. (b) Rabi revival in a 13.2-photon coherent field. Dots with statistical error bars joined by a thin black line: experimental probability $P_g(t_e)$. Solid red line: numerical simulation of the experiment. The inset shows a zoom on the revival. (c) Fourier transform of the signal in (b) (black dots). The solid red line is a fit to a superposition of Gaussian peaks. The vertical dotted blue lines are at the expected Rabi frequencies for photon numbers, $n$, given on top. The blue solid line is the envelope of the Poisson photon number distribution for a 13.2 photon coherent state, adjusted in height to fit the abitrary scale of the Fourier Transform.}
\label{fig:rabizero} 
\end{figure}

Figure~\ref{fig:rabizero}(a) shows the experimental $P_g(t_e)$ (dots with statistical error bars, joined by a thin black line) as a function of the effective interaction time $t_e$ taking into account the motion of the atom through the Gaussian mode geometry~\supp. We observe nearly 20 vacuum Rabi periods, a considerable improvement over previous experiments~\cite{BruneQuantumRabiOscillation1996}. The red line in Fig.~\ref{fig:rabizero}(a) results from a numerical simulation of the experiment taking into account the detection imperfections, resulting in a reduced oscillation contrast, as well as the residual initial thermal field and cavity relaxation towards thermal equilibrium~\supp. The Rabi frequency,  $\Omega_0/2\pi=49.8$~kHz, the probability $p_1=0.09$ for having initially one photon in $C$ and the precise position of the exciting laser beams inside the cavity mode, defining the relation between the real and effective interaction times are extracted from a fitting procedure described in~\supp.

We record now the Rabi oscillation in a coherent state $\ket\beta$ with $|\beta|^2=\overline{n}\approx 13.2\pm 0.1$ [timing in Fig.~\ref{fig:scheme}(b), with $t_\beta=14\ \mu$s]. Note that $\beta$ can be assumed to be real without loss of generality. The signal is plotted in Fig.~\ref{fig:rabizero}(b) (points with statistical error bars joined by a thin black line). It clearly exhibits the first revival around $T_r= 146\ \mu$s. The signal is in excellent agreement with a numerical simulation of the experiment (red solid line) taking into account the same imperfections as those in Fig.~\ref{fig:rabizero}(a).

The Fourier transform of the revival signal [dots in Fig.~\ref{fig:rabizero}(c) with a fit to a sum of gaussians -- solid red line] exhibits discrete peaks, very close to the expected Rabi frequencies in an $n$-photon field, $\Omega_0\sqrt{n+1}/2\pi$ (vertical blue lines). They are well-resolved up to about 18 photons. This signal provides a textbook evidence of field quantization. The weight of the peaks directly measures the photon number distribution,  $p(n)$~\supp. It is in excellent agreement with the expected  Poisson distribution  for a mean initial photon number $13.2$ (blue solid line). 

Close to half the revival time, $t_e\approx T_r/2\approx 73\ \mu$s, the atomic state is expected to factor out~\supp. The cavity is then in a superposition of two components with nearly opposite phases~\cite{Gea-BanaclocheCollapserevivalstate1990,Gea-BanaclocheAtomfieldevolution1991}. The photon number parity ${\cal P}=\sum_n(-1)^np(n)$ oscillates with $t_e$~\cite{BirrittellaPhotonnumberparityoscillations2015} and the field Wigner function, $W(\alpha)$, exhibits fringes near the origin of phase space revealing the cat's coherence~\cite{DelegliseReconstructionnonclassicalcavity2008}. 

In order to probe the cavity state, we make use of its resonant interaction with a `probe' atom, initially in $\ket e$. For a well-chosen~\supp\ effective interaction time $t'_e$ close to $T_r/2$, the probability $P_g$ for finding the atom in $\ket g$  is  $P_g(t'_e)\approx\left(  1- {\cal P} \right)/2$, and $P_g$ measures the parity ${\cal P}=\pi W(0)/2$. By displacing the cat state by $\alpha$ before the resonant interaction, we get access to $W(-\alpha)$.

The timing of the experiment is depicted in Fig.~\ref{fig:scheme}(c). We set the first resonant effective interaction time to $t_i=60\ \mu$s. After the atom has been tuned out of resonance, we perform a displacement by a real amplitude $\alpha$ (injection time $t_\alpha$). Simultaneously, we get rid of  the $\ket g$ part of the atomic state by a resonant radio-frequency pulse transferring $\ket g$ to low-$m$ states, finally undetected. Thus, any detected atom was in $\ket e$ at the beginning of a new resonant interaction, starting after a delay $t_d= 6\ \mu$s and lasting for a variable time $t'_i$ corresponding to the effective time $t'_e$.

\begin{figure}
\includegraphics[width=8.5cm]{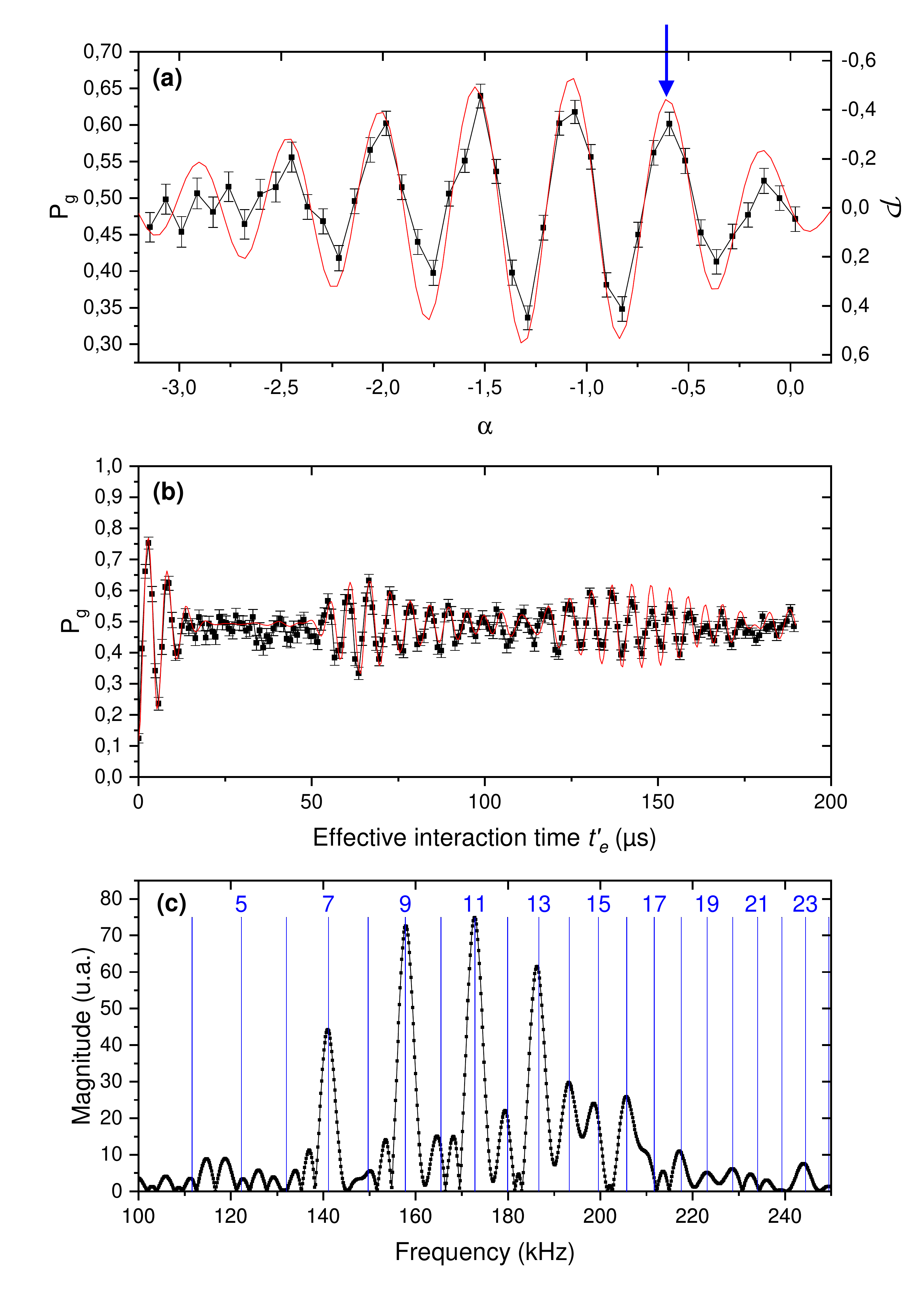}
\caption{(a) Measurement of the displaced cat's parity proportional to its Wigner function. $P_g(t'_e,\alpha)$ as a  function of $\alpha$ for $t'_e=68.5\ \mu$s (dots with statistical error bars joined by a thin solid black line). The solid red line results from a numerical simulation. (b) Quantum revivals in a displaced cat state. $P_g(t'_e,\alpha)$ as a function of $t'_e$ for $\alpha=-0.60$ (dots with statistical error bars and a connecting thin solid black line). The solid red line results from a numerical simulation. The additional revival around $T_r/2$ is clearly visible. (c) Fourier transform of the signal in (b) (dots with connecting thin black line). Vertical blue lines depict the expected Rabi frequencies. The absence of even photon numbers in the distribution is conspicuous.}
\label{fig:halfcat} 
\end{figure}

Figure~\ref{fig:halfcat}(a) shows $P_g(t'_e,\alpha)$  (dots with statistical error bars) as a function of $\alpha$ for $t'_e=68.5\ \mu$s $\simeq T_r/2$. We have chosen this $t'_e$ value to optimize the contrast of the oscillations, so that $P_g(t'_e,\alpha)\approx[1- \pi W(-\alpha)/2]/2 $. The solid red line results from the simulation of the experiment and is in excellent agreement with the data. From the period of the oscillations in Fig.~\ref{fig:halfcat}(a), we deduce the size of the cat, $D^2$,  measured by the square of the distance $D$ of the superposed components in phase space~\supp. We get $D^2=45.1\pm 0.4$ photons, in excellent agreement with the value provided by the numerical simulation. This size compares with that of the largest cats ever prepared by dispersive interaction~\cite{VlastakisDeterministicallyEncodingQuantum2013}.

Figure~\ref{fig:halfcat}(b)  shows $P_g(t'_e,\alpha)$ as a function of  the effective interaction time $t'_e$ for $\alpha=-0.60$. This displacement [arrow on Fig.~\ref{fig:halfcat}(a)]  brings one of the extrema of $W$ onto the origin of phase space, resulting in a displaced cat with a maximum parity. We clearly observe here a dual revival, one at the standard revival time $T_r$ and one earlier, at $T_r/2$, revealing the displaced cat  parity. The agreement of the experimental data with the simulation (red solid line) is excellent.

The Fourier transform of the Rabi signal, shown in Fig.~\ref{fig:halfcat}(c), conspicuously displays the displaced cat parity. The photon number distribution inferred from this spectrum provides a displaced cat parity ${\cal P}=-0.48$. The simulation provides ${\cal P}=-0.41$ in excellent agreement with the observed value. The parity for an ideal experiment taking into account cavity relaxation, is expected to be  $-0.49$, very close to the observed one. We thus generate large cats with a good fidelity.

\begin{figure}
\includegraphics[width=8.5cm]{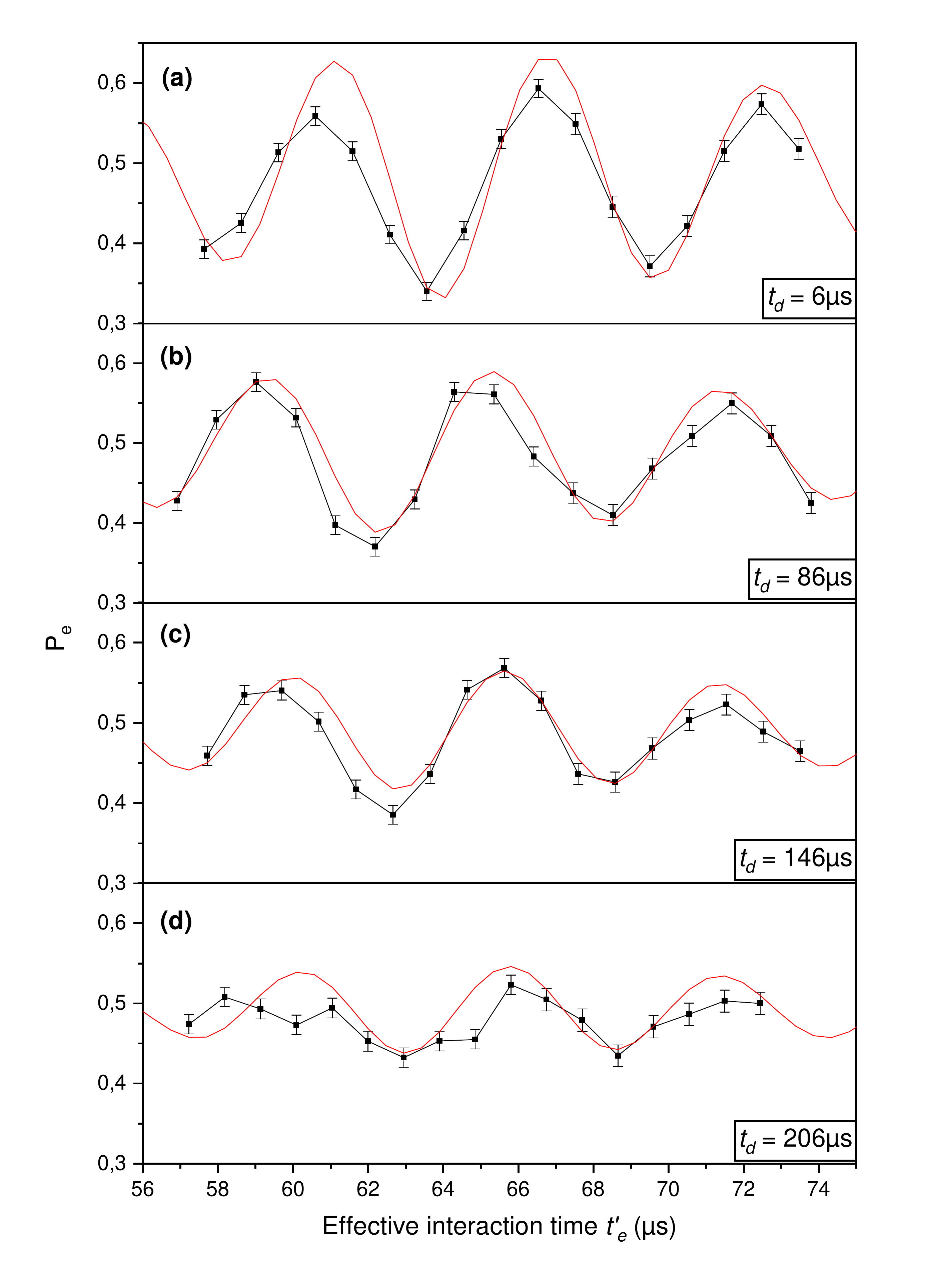}
\caption{(a)-(d) Rabi oscillations signals around $T_r/2$ for the probe atom in a displaced cat state for $\alpha=-0.60$ and four delay times $t_d=6$, $86$, $146$ and $206\ \mu$s. Dots with statistical error bars and connecting thin line are the experimental data. The solid red line results from the simulation of the experiment. The fast decrease of the oscillation contrast reveals the cat decoherence.}
\label{fig:decoh} 
\end{figure}

The decoherence time scale of the cat in the finite temperature environment of the cavity is expected to be $ 200$~$\mu$s~\cite{Supplementarymaterial,KimSchrodingercatstatesfinite1992a}. In order to investigate this fast decoherence, we monitor the amplitude of the revival at $T_r/2$ as a function of the delay time $t_d$ (note that the displacement by $\alpha$ does not change the dynamics of decoherence).

Fig.~\ref{fig:decoh}(a-d) shows four Rabi oscillations signals for $t'_e$ close to $T_r/2$ for $\alpha=-0.60$ and for four values of $t_d$ (dots with statistical error bars and simulation of the experiment as a solid red line). The agreement between simulation and data is excellent, confirming the decoherence time scale. Note that the moderate $T_c$ value, much shorter than 130~ms reached in previous experiments ~\cite{DelegliseReconstructionnonclassicalcavity2008}, is appropriate for an observation of the cat's decoherence in the available interaction time range.  

In conclusion, we have observed the resonant Rabi oscillation in a coherent field in an unprecedented range of photon numbers. We have shown the generation of large cat states through the first observation of an early revival for the Rabi oscillation in a cat with a well-defined parity. We have monitored the fast decoherence of these non-classical states. 

The resonant interaction generates cat states efficiently and significantly faster than the dispersive method~\supp. Using more than one atom, it can lead to the preparation of more complex state superpositions with multiple components~\cite{PathakGenerationsuperpositionmultiple2005,MeunierEntanglementdecoherenceatoms2006}. It is thus a promising method for fundamental decoherence studies, but also,  in the cQED context, for the use of cats in quantum error correction protocols~\cite{OfekExtendinglifetimequantum2016,RosenblumFaulttolerantdetectionquantum2018}.
 
This experiment also opens the way to a new realm for atomic physics CQED, with extremely long interaction times and extremely low damping cavities. Particularly promising is the use of laser-trapped circular Rydberg atoms~\cite{NguyenQuantumSimulationCircular2018}. They could be combined with a 3D-microwave structure sustaining a high-quality resonant mode, but admitting no other modes resonant with the atomic spontaneous emission decay channels. One might then combine the best of the CQED and cQED worlds, with reproducible atoms, well-controlled and understood decoherence channels, nearly infinite interaction times and the slow pace of CQED experiments instrumental for real-time control~\cite{SayrinRealtimequantumfeedback2011}.

\acknowledgements{We acknowledge funding by the EU under the ERC projet `DECLIC' (Project ID: 246932) and the RIA project `RYSQ' (Project ID: 640378).}

\renewcommand{\theequation}{S\arabic{equation}}
\renewcommand{\thefigure}{S\arabic{figure}}
\setcounter{figure}{0}

\clearpage
\section{Supplementary material: Quantum Rabi oscillations in coherent and in mesoscopic Schr\"odinger ``cat" field states}

We first give a simple theoretical model of the experiment. We then provide additional details about the experimental apparatus and the data acquisition. We then discuss the relations between the real resonant interaction time and the effective one. Finally, we discuss the data processing and the numerical simulation of the experiment.

\subsection{Collapse, revival and cat generation: theoretical remarks}

We examine here the collapse and revival for an initial coherent  or cat-like state. We consider an atom initially in $\ket e$. The cavity $C$ contains a quantum state $\Psi=\sum_nc_n\ket n$ ($\ket n$ is the Fock state with $n$ photons), corresponding to the photon number distribution $p(n)=|c_n|^2$. For a coherent state, $p(n)=p_c(n)$, with $p_c(n)= e^{-\overline n}{\overline n}^n/n!$, where $\overline n$ is the average photon number. A simple calculation~\cite{HarocheExploringquantumatoms2006} provides the probability $P_g(t)$ for detecting the atom in $\ket g$ after an effective interaction time $t_e$
\begin{equation}
P_g(t_e)=\frac{1}{2}\left[  1-\sum_n p(n)\cos\left( \Omega_0\sqrt{n+1}\, t_e  \right)  \right]\  .
\label{eq:rabi}
\end{equation}
The Rabi signal is a sum of the Rabi oscillations in the Fock states $\ket n$, at a frequency $ \Omega_0\sqrt{n+1}$, weighted by the photon number distribution. The complex interplay of the non-rational frequencies appearing in this sum leads to the collapse and revival phenomenon.

Let us first consider the case of a coherent state, $\ket{\beta}$, with a $\sqrt{\overline n}$ dispersion of the photon number distribution, $p_c(n)$, around $\overline n$. For short effective interaction times, the dispersion of the oscillation phases in Eq.~(\ref{eq:rabi}) is negligible and we find, as expected, a Rabi oscillation at the average angular frequency $\Omega_r=\Omega_0\sqrt{\overline n}$ with a close-to-unity contrast. The collapse is due to the spread of the Rabi frequencies in Eq.~(\ref{eq:rabi}). In order to estimate the collapse time, we can compute the time, $T_c\simeq 2\pi/\Omega_0$, after which the Rabi oscillations for two photon numbers separated by the photon number variance, $n=\overline n \pm\sqrt{\overline n }/2$, get in phase opposition. A more detailed calculation shows that the envelope of the oscillations collapse is a gaussian with a $1/e$ time-width $T_c= 2\sqrt 2/\Omega_0$~ \cite{HarocheExploringquantumatoms2006}.

The revivals correspond to the rephasing of the Rabi oscillations associated to different photon numbers. For $n\simeq\overline n$, it is easy to see that the oscillations for $n$ and $n+1$ photons come back in phase at a time $T_r=4\pi\sqrt{\overline n}/\Omega_0$. Since this time is independent of the chosen $n$ value, all oscillations come back in phase at that time, resulting in the maximum of the revival.

An early quantum revival can be expected near $t_e=T_r/2$ if the initial cavity state $\ket\Psi$ has a well-defined photon number parity ${\cal P}=\sum_n (-1)^n p(n)=\langle \exp(i\pi N)\rangle$ ($N$ is the photon number operator). This is the case for an ideal `phase ' or `parity cat'  $(\ket{\gamma}\pm\ket{-\gamma})/\sqrt 2$ (we assume $\gamma\gg 1$ for the normalization factor), for which ${\cal P}=\pm 1$. 

Choosing $t_e= 2\pi\sqrt{n_0}/\Omega_0+\tau$, where $n_0$ is an even integer close to $\overline n$, we get from Eq. (\ref{eq:rabi}), expanding $\sqrt{n+1}$ at the first order in $1/\sqrt{ n_0}$
\begin{equation}
P_g(t_e)=\frac{1}{2}\left[  1- {\cal P} \cos\Phi  \right]\  ,
\end{equation}
where
\begin{equation}
\Phi\approx\Omega_r \tau.
\end{equation}
We thus get Rabi oscillations at the frequency $\Omega_r$, with an amplitude proportional to the parity $\cal P$ of the initial state. For an initial coherent state, ${\cal P}=\exp(-2\overline n)$ and there is no sizable Rabi oscillation near $T_r/2$. For a parity cat state, there are instead well-contrasted oscillations at half the revival time. This result can be retrieved by a simple argument. For a parity-cat, $p(n)$ cancels for odd or even photon numbers. The least spacing between Fock states contributing to the sum in Eq.~(\ref{eq:rabi}) is 2 and the rephasing time is thus reduced by a factor of 2.

Choosing a time $\tau$ such that  $\cos\Phi =1$, the measurement of $P_g$ directly provides $\cal P$, proportional to the field's Wigner function at the origin, $W(0)$~\cite{LutterbachMethodDirectMeasurement1997}.  Adding a translation of the initial state by an amplitude $\alpha$ [represented by the displacement operator $D(\alpha)$] before starting the Rabi oscillation thus directly determines the Wigner function, $W(-\alpha)$ at one point in phase space:
\begin{equation}
W(-\alpha)=\frac{2}{\pi}\langle D(-\alpha) \cal P D(\alpha) \rangle\ .
\end{equation}

This simple analysis in terms of a beat between Rabi oscillations leaves aside the evolution of the field state. A much more detailed insight is provided in~ \cite{Gea-BanaclocheCollapserevivalstate1990,Gea-BanaclocheAtomfieldevolution1991,BuzekSchrodingerCatStatesResonant1992,HarocheExploringquantumatoms2006}. Since the initial field injection in $C$ provides the phase reference, we can assume that $\beta$ is real and positive without loss of generality. Expanding the full atom-cavity state to the first non-trivial order in $(n-\overline{n})/\overline{n}\ll 1$, we find that the atom-field state $\ket{\Psi(t_e)}$ at the effective interaction time $t_e$ reads
\begin{eqnarray}
\ket{\Psi(t_e)}=\frac{1}{\sqrt 2}&&[\ket{\psi_a^+(t_e)}\otimes\ket{\psi_c^+(t_e)}\nonumber\\
&+&\ket{\psi_a^-(t_e)}\otimes\ket{\psi_c^-(t_e)}]\ ,
\end{eqnarray}
where 
\begin{equation}
\ket{\psi_a^\pm(t_e)}=\frac{1}{\sqrt 2}e^{\pm i\Omega_r t_e/2}\left[ e^{\pm i\Omega t_e}\ket e\mp \ket g \right]\ ,
\end{equation}
\begin{equation}
\ket{\psi_c^\pm(t_e)}=e^{\mp i\Omega_rt_e/4}\ket{\beta e^{\pm i\Omega t_e}}\ ,
\end{equation}
and $\Omega=\Omega_0/4\sqrt{\overline n}$.

The state $\ket{\Psi(t_e)}$ is generally an entangled atom-field state, a cat state in which two atomic states $\ket{\psi_a^\pm(t_e)}$, slowly rotating at frequency $\pm\Omega$ in the equatorial plane of the Bloch sphere, are correlated to two coherent field components, also rotating at  $\pm\Omega$ in the Fresnel plane from the initial $\ket\beta$ state. The two coherent components overlap for $t_e$ close to 0 and for $t_e$ close to all multiples of $T_r$. The field state then factors out and the high-contrast Rabi oscillations  at $\Omega_r$ result from a quantum beat between the two atomic states. In this simple picture, the revivals are periodic. In fact, the higher order terms in the developments in powers of  $(n-\overline{n})/\overline{n}$ result in a progressive phase spread (squeezing) of the coherent components. The long-term revivals are thus longer and less contrasted than the first. Already at the first revival, the phase spread is of the same order as the initial phase uncertainty in $\ket\beta$~ \cite{HarocheExploringquantumatoms2006}. It is thus about twice longer than the collapse and has a maximal ideal contrast close to 50\%. 

Another interesting situation occurs at half the revival time, $T_r/2$, for which $\Omega t_e=\pi/2$. The two atomic states are thus both equal, within a global phase, to $(\ket e-\ket g)\sqrt 2$. The atom disentangles from the field. The probability $P_g$ for detecting it in $g$ is 1/2 and the field is then left in 
\begin{equation}
\ket{\psi_{cat}}=e^{i\pi\overline n}\ket{i\beta}-\ket{-i\beta}\ ,
\end{equation}
a cat state with the parity
\begin{equation}
{\cal P}_{cat}=-\cos(\pi\overline n)\  .
\end{equation}

Let us compare this resonant generation of a cat state with the dispersive one, commonly used in Cavity Quantum Electrodynamics~\cite{DelegliseReconstructionnonclassicalcavity2008} and particularly in circuit QED~\cite{VlastakisDeterministicallyEncodingQuantum2013}. In the dispersive regime, the atom in state $\ket e$ or $\ket g$ causes a shift of the cavity frequency by $\pm\Omega_0^2/4\delta$ ($\delta$ is the atom-field detuning). Preparing the atom in a state in the equatorial plane of the Bloch sphere and detecting it after a $\pi/2$ resonant pulse mixing again $\ket e$  and $\ket g$ after the interaction with the cavity (and thus erasing any which-path information about the atomic state in the cavity), we prepare a cat state, superposition of the coherent components $\ket{\beta\exp(\pm i\varphi)}$, with $\varphi=\Omega_0^2t_d/4\delta$ ($t_d$ is the effective interaction time for a dispersive interaction, close to the effective interaction time for a resonant interaction, $t_e$).  A parity cat is reached for $t_d=2\pi\delta/\Omega_0^2$. When $\delta>\Omega_0\sqrt{\overline n}$, a condition essential to enforce the dispersive regime, the dispersive interaction is slower by a factor $\delta/\Omega_0\sqrt{\overline n}$ than the resonant one to generate a phase cat. A more detailed discussion, taking into account intermediate regimes, shows that the cat preparation time is minimal, all other things equal, when $\delta=0$~\cite{HarocheExploringquantumatoms2006}.

The cat states are very sensitive to decoherence. The Lindblad equation formalism leads to an exact expression for the density matrix of a relaxing cat state. In an environment at zero temperature, the coherence between the cats components is found to decrease, for $t\ll T_{Cav}$, as $\exp(-2\overline n t/T_{Cav})$~\cite{HarocheExploringquantumatoms2006}. The decoherence time scale is thus $T_D=T_{Cav}/(2\overline n)$, $2\overline n$ times shorter than the cavity energy damping time. This fast relaxation is a key feature of the decoherence of mesoscopic state superpositions. It can be qualitatively understood in terms of parity. Starting from an odd cat, a photon loss leads to an even cat. After having lost on the average about one photon, the field is in a mixture of odd and even cats, which is also a statistical mixture of the two coherent components. The coherence is thus lost after about one photon has disappeared, after a time of the order of $T_{Cav}/\overline n$. At a non-zero temperature, the decoherence time is $T_D=T_{Cav}/2[\overline n (1+2n_{th})+n_{th}]$, where $n_{th}$ is the average thermal photon number~\cite{KimSchrodingercatstatesfinite1992a}.

\section{Experimental details}

\begin{figure}
\includegraphics[width=8.5cm]{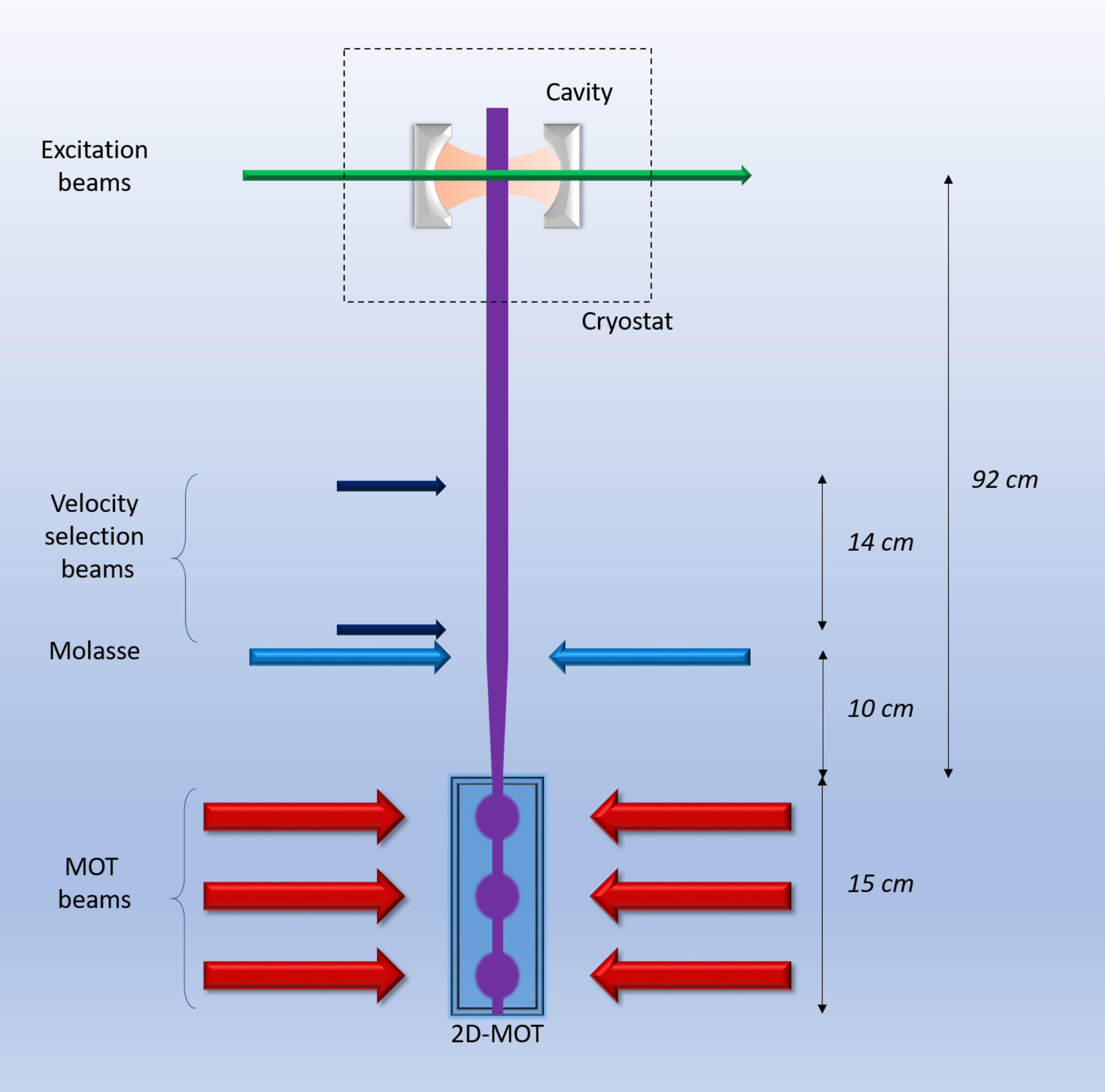}
\caption{Detailed scheme of the experiment with all beams and an indication of distances}
\label{fig:detailed} 
\end{figure}

\begin{figure}
\includegraphics[width=8.5cm]{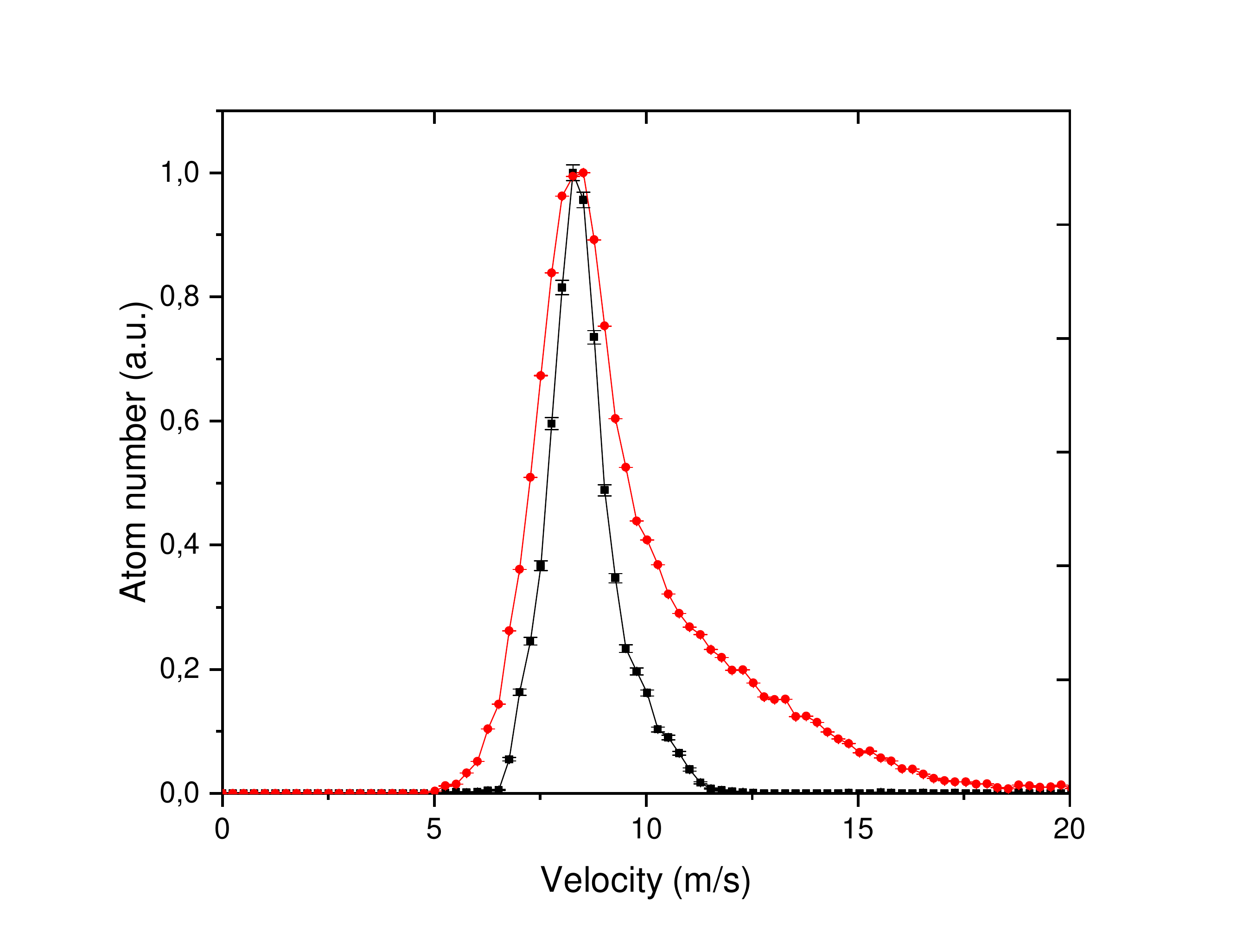}
\caption{Velocity distributions of the atomic beam, with (black dots with connecting thin black line) and without (red dots with connecting line) strobsoscopic selection. The red curve is renormalized to have a comparable scale to the other one. The errors bars represent the statistical dispersion.}
\label{fig:velocity} 
\end{figure}

A detailed scheme of the experiment is shown in Fig.~\ref{fig:detailed}. A slow vertical $^{85}$Rubidium atomic beam effuses from a 2D-MOT (SYRTE Laboratory). Its transverse velocity spread is reduced by a transverse optical molasses stage, situated right above the 2D-MOT arrangement. The longitudinal beam velocity distribution, shown in Fig.~\ref{fig:velocity} as a red line, is inferred from a circular-atom time-of-flight measurement  from the laser-excitation region, close to the center of the cavity, to the field-ionization detector (note that the velocity-dependent efficiency of the circular state preparation and the finite lifetime of the circular levels slightly distort the measured distribution). The distribution is peaked at about 8.50~m/s, with a width of 2.4~m/s. This dispersion is too large for a good control of the experimental sequence.

We therefore perform an additional  passive longitudinal velocity-selection stage between the molasses and the cryostat. Two resonant laser beams, separated by 14~cm are used to push the atoms out of the beam axis, interrupting the atomic flux. They are periodically interrupted for 5~ms each 15~ms. The atoms reaching the cavity must have passed the two selection beams while they were off. The resulting velocity distribution is shown as a black line in Fig.~\ref{fig:velocity}. It is centered at 8.36~m/s, with a width 1.2~m/s. 

The circular atoms are prepared by laser, rf and mw excitation inside the cavity $C$~\cite{SignolesCoherentTransferLowAngularMomentum2017}. We use three lasers, perpendicular to the atomic beam at 780, 776 and 1258~nm resonant on the $5S_{1/2}\rightarrow5P_{3/2}$, $5P_{3/2}\rightarrow5D_{5/2}$ and $5D_{5/2}\rightarrow 52F$ transitions. The pulsed laser excitation (0.1~$\mu$s pulse for the third laser) is performed in a weak electric field $F$ (0.14~V/cm) produced by a voltage applied across the cavity mirrors. The Stark structure being resolved, we only excite levels with a magnetic quantum number $m=\pm 2$. We then raise $F$ to 2.4~V/cm. The atomic state adiabatically evolves into  the lowest $m=\pm2$ levels in the hydrogenic Stark manifold. We apply a $\sigma^+$-polarized rf field at 230~MHz, slightly below resonance with the Stark frequency in the manifold, while slowly decreasing $F$. The $m=2$ level is efficiently transferred to the circular state $52C$ by an adiabatic rapid passage process. It leaves the $m=-2$ state unaffected (transitions to the $m=-1$ Stark levels are out of resonance due to the finite quantum defects of the low-$\ell$ states). These $m=-2$ states do not interact with $C$ and are not finally detected.  A final $1\ \mu$s microwave pulse  prepares the initial level $51C$ ($\ket e$), leaving in the 52 manifold spurious high-$m$ states and about 20~\% of the 52C population. 

For initial field erasing, we send resonant absorbing atoms in $C$ before the start of the sequence. We use a two-photon mw pulse resonant on the $52C\rightarrow 50C$ to prepare directly about 10 atoms in $\ket g$. The last field-erasing atoms are prepared 2.7~ms before the the start of the actual sequence so that they do not contaminate the field-ionization detection of the useful samples.

The cavity $C$ is made up of Niobium-coated diamond-machined copper mirrors. We measure its resonant frequency and its lifetime (8.1~ms) by injecting a field into it via its diffraction loss channels. Probe atoms are used to measure this field and its decay versus time~\cite{KuhrUltrahighfinesseFabryPerot2007}.

The field injections  in the experimental sequence are also performed by $S$, used with a constant power. The injected amplitude modulus $|\beta|$, measured at time $t=0$ in the sequence (the time at which the atom is first set at resonance with the mode) is a linear function of the injection time $t_\beta$. We obtain an approximate calibration of  this linear law by recording the first periods of the resonant Rabi oscillation as a function of the injection time and by fitting them to a simulation. A better determination is provided by the Rabi revival signal itself, whose Fourier transform directly measures the photon number distribution. A slight refinement within the error bars of the previous method is provided by adjusting finely the scale of the amplitude axis used for the experimental data in Fig.~3(a) (main text). We finally use the linear law given in the main text. The small offset on the injection time is due to the finite commutation time of the X-band source.  

The  relative phase, $\varphi$, of the two injections used for the parity measurements is set to $\pi$ by adjusting the detuning $\Delta$ of the source  with respect to the cavity (in the cavity-field rotating frame, the source field rotates at $\Delta$). The 103.8~$\mu$s time separation between the two injections makes $\varphi$ more sensitive to the precise value of the cavity frequency than the Rabi oscillation itself (a cavity shift of only 2.4~kHz, much lower than $\Omega_0/2\pi$, shifts $\varphi$ by $\pi$). We thus check the cavity resonance frequency every day. We have devoted considerable efforts to the reduction of cavity mechanical vibrations due, in particular, to cryogenic fluids oscillations.

The field-ionization detector $D$ is active between 7.5 and 9~ms after the initial laser pulse, so that we detect the lower-velocity part of the beam distribution, providing an additional velocity-selection stage. During their time of flight towards $D$, the spontaneous decay of the circular states cannot be neglected (lifetime of the order of 30~ms). A part of the atoms initially in $\ket e$ is thus transferred into $\ket g$ and affects the measured population. Similarly, a part of the atoms left in the 52 manifold may decay in the 51 one and be spuriously detected in $\ket e$. In order to avoid cross-talks between levels populations, we first transfer, 450~$\mu$s after the laser excitation,  the atoms in $\ket g$ into the $48C$ circular state  with a two-photon microwave pulse. We then use, 600~$\mu$s after the laser pulse, a resonant rf pulse, transferring the atoms of the 52C manifold into low-$m$ states, which have a ionization threshold very different from that of the circular state (and which, moreover, rapidly decay by optical transitions). A similar procedure is used to get rid of the atoms left in $\ket g$ after the first Rabi oscillation period in the cat parity measurements.

\section{Data analyzis}

\begin{figure}
\includegraphics[width=8.5cm]{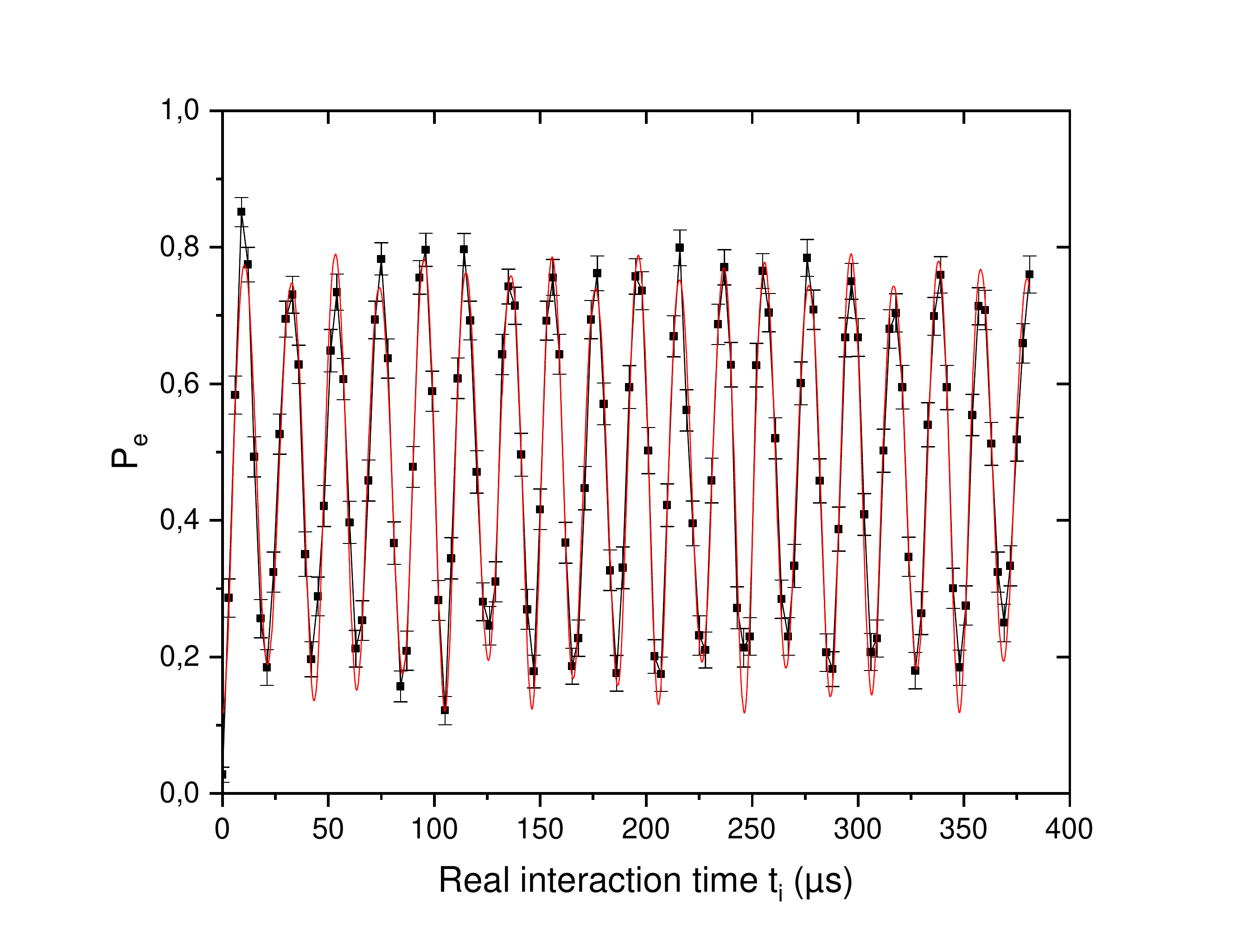}
\caption{Vacuum Rabi oscillation. Dots with statistical error bars joined by a thin black line: experimental probability $P_g(t_i)$ as a function of the real interaction time $t_i$. Solid red line: fit on the theoretical model we use to get all unknown parameters.}
\label{fig:rerabi} 
\end{figure}

The essential parameters of the experiment are extracted from a direct fit of the raw vacuum Rabi oscillation signal, $P_g(t_i)$, recorded as a function of the interaction time $t_i$. This signal is shown in Fig.~\ref{fig:rerabi} (note that the signal in Fig.~2(a) of the main text is plotted as a function of the effective interaction time $t_e$). The fitting function is the expected Rabi oscillation in a mixture of zero and one-photon states describing the initial residual thermal field in the cavity. The ideal $P_g$ values,  $P^i_g$, are then given by 
\begin{equation}
P^i_g=(1-p_1)\cos^2(\Omega_0 t_e/2)+p_1\cos^2(\sqrt{2}\Omega_0 t_e/2)
\label{eq:rabiideal}
\end{equation}
where $p_1$ is the initial probability for having one photon in $C$. The time variable in this expression is the effective interaction time $t_e$. It takes into account the motion of the atom along axis $Ox$ through the gaussian transverse electromagnetic TEM mode geometry, characterized by its waist $w=6$~mm. It reads, as a function of $t_i$:
\begin{equation}
t_{e}(t_i)=\frac{w}{v}\frac{\sqrt{\pi}}{2}\left(\erf\left((t_i-x_0 v)\frac{v}{w}\right)-\erf\left(-x_0 v\frac{v}{w}\right)\right),
\end{equation}
In this expression, $x_0$ is the initial ($t=0$) position of the atom on the $Ox$ axis, determined by the position of the exciting laser beams. It is not precisely known and must be extracted from the fitting procedure. The atomic velocity $v=8.1\pm 0.1$~m/s is well-known from the velocity distribution shown in Fig.~\ref{fig:velocity}.

We then have to take into account in the fit the detection errors. They include the $50C-48C$ probe efficiency, the efficiency of the rf pulse used to get rid of the $52C$, the contamination of $51C$ into $50C$ due to spontaneous emission between cavity interaction and the pulse transferring $50C$ into $48C$, the contamination of $50C$ into $51C$ due to absorption of thermal photons and the detection efficiency difference between the 51 and 48 levels. All these errors can be encapsulated in a simple homographic law transforming the ideal $P^i_g$ signal into the measured one, $P_g$. For $N$ repetitions of the experiment, the number of atoms actually detected in $e$ or $g$, $N_{e,g}^d$ can be written as a function of the ideal counts, $N_{e,g}^i$, as
\begin{equation}
N_{e,g}^d=A_{e,g}N_{e,g}^i+B_{e,g}N_{g,e}^i+C_{e,g}\ .
\end{equation}
Using $N_{g}^i+N_{e}^i=N$ and after simple algebra, we get
\begin{equation}
P_g=\frac{N^d_g}{N^d_g+N^e_g}=\frac{aP^i_g+b }{cP^i_g+d }\ ,
\label{eq:homo}
\end{equation}
where
\begin{eqnarray}
a=A_g-B_g&\ ,\quad &c=A_g-B_g+A_e-B_e \nonumber\\
b=B_g+C_g&\ ,\quad &d=B_g+C_g+B_e+C_e
\end{eqnarray}
We use the homographic law of Eq.~(\ref{eq:homo}) to transform the ideal Rabi oscillation of Eq.~(\ref{eq:rabiideal}) and we fit the resulting signal onto the raw Rabi oscillation. The result of the fit is shown as a red solid line in Fig.~\ref{fig:rerabi}. We extract from this fit all unknown parameters, namely
\begin{eqnarray}
\Omega_0&=&2\pi\times (49.88\pm 0.03) \textrm{~kHz}, \nonumber\\ x_0&=&1.72\pm 0.02\textrm{~mm},\nonumber\\
\quad p_1&=&0.094\pm 0.014 \nonumber\\
a&=&1, \nonumber\\ b&=&0.133\pm 0.02, \nonumber\\ c&=&0.297\pm 0.07, \nonumber\\ d&=&1.136\pm 0.08
\end{eqnarray}

The $\Omega_0$ value is very close to the expectation based on the cavity geometry ($2\pi\times 50$~kHz). The slight discrepancy may be due to a slight offset of the laser beam with respect to the center of the cavity in the $Oz$ direction. The $p_1$ value is in fair agreement with the expectations of an ideal cavity erasure procedure, followed by a return to thermal equilibrium (0.34 photons on the average) for 2.7~ms. The $a,b,c,d$ values are also in qualitative agreement with independently measured detection errors.

We use the same set of parameters for all numerical simulations shown as red lines in the figures of the main text. These  simulations are performed with the Qutip library~\cite{JohanssonQuTiPPythonframework2013}, computing the interaction hamiltonian between a 2-level atom and the harmonic oscillator of the mode of the cavity, truncated to the first 50 Fock states. We take into account the finite lifetime of the cavity and of the atom, the original thermal state we start from, and the field relaxation towards the steady thermal state of the experiment. 

We use Fourier transform to convert the Rabi oscillation data to the frequency domain and get straight vision of the photon number distribution. In oder to do that and obtain sharp peaks the experimental signal is first linearly interpolated with a 0.1~$\mu$s step. We then substract its mean value, symmetrize it with respect to $t=0$, and complete the data with 0 until $t=\pm3000~\mu$s. We finally compute the fast Fourier transform with a rectangle window.


\end{document}